# Influence of Si doping and O$_2$-flow on arc deposited (Al,Cr)$_2$O$_3$ coatings

L. Landälv [1,2,a)], E. Göthelid [2], J. Jensen [1], G. Greczynski [1], J. Lu [1], M. Ahlgren [1], L. Hultman [1], B. Alling [3], P. Eklund [1,a)]

[1] Thin Film Physics Division, Department of Physics, Chemistry, and Biology (IFM), Linköping University, Linköping SE-581 83, Sweden
[2] Sandvik Coromant AB, Stockholm SE-126 80, Sweden
[3] Theoretical Physics, Department of Physics, Chemistry, and Biology (IFM), Linköping University, Linköping SE-581 83, Sweden

[a)] Corresponding authors: email address: ludvig.landalv@sandvik.com and Per.eklund@liu.se

**Abstract**

(Al,Cr)$_2$O$_3$ coatings with Al/(Al+Cr) = 0.5 or Al = 70 at.%, doped with 0, 5 or 10 at.% Si, were deposited on hard metal and Si(100) substrates to elucidate the influence of Si on the resulting coatings. The chemical analysis of the coatings showed between 3.3 and 7.4 at.% metal fraction Si incorporated into all studied coatings depending on cathode Si-composition. The incorporated Si content does not change significantly with different oxygen flow covering a wide range of deposition conditions from low to high O$_2$ flow during growth. The addition of Si promotes the metastable B1-like cubic structure over the thermodynamically stable corundum structure. The hardness determined by nanoindentation of the as-deposited coatings is slightly reduced upon Si-incorporation as well as upon increased Al-content. Si is found enriched in droplets but can also be found at a lower content, evenly spread, without visible segregation on the ~5 nm scale, in the actual oxide coating. The positive effect of improved cathode erosion upon Si-incorporation has to be balanced against the promotion of the metastable B1-like structure, having lower room temperature hardness and inferior thermal stability compared to the corundum structure.

# I. INTRODUCTION

Aluminum oxide exists in several crystallographic structures, which are used for different purposes. For example γ-$Al_2O_3$ has been used as catalysis material [1,2] and κ-$Al_2O_3$ has been applied for metal cutting applications, i.e. like milling of stainless steel [3]. Amorphous and corundum-structure aluminum oxide both have important use as dielectric material for microelectronics and GHz data communication [4-8]. The thermodynamically stable phase is corundum, with a rhombohedral crystal structure (α-$Al_2O_3$, space group R3c [9]), and has been widely used within the cutting tool industry since the beginning of the 1990's (first use in the 1970's) [3]. This is due to its high melting point and chemical stability with respect to the workpiece material, combined with low thermal conductivity, properties summarized in [10,11]. It also possesses high hot hardness [10,12] and low adhesion to the workpiece material [13,14]. Chemical vapor deposition (CVD) has been used to produce thick and uniform α-$Al_2O_3$ coatings [15]. However, the high deposition temperature inherent to the process leads to high thermally induced tensile stresses upon cooling. Since the thermal expansion coefficient is larger for aluminum oxide than for the hard metal substrate (WC-Co), the coating tends to obtain tensile stresses and cracks forms upon cooling from the high deposition temperature 1000-1050 °C [10,15-17].

α-$Al_2O_3$ made by physical vapor deposition (PVD) has been a long-sought goal in coating development since it would permit to produce compressive stresses in the coating, due to ion bombardment, combined with a lower deposition temperature as PVD operates far from thermal equilibrium. The lower depositions temperature for the PVD process compared to the α-$Al_2O_3$ CVD process would also reduce the tensile stresses in the coating caused by the difference in thermal expansion between substrate and coating [15]. Having a lower deposition temperature would also be beneficial for being able to use other types of substrates, like high-speed steel, which is more heat sensitive [18,19]. Cobalt diffusion from the substrate into the coating can also be an issue at CVD-deposition temperatures. This phenomena has also been encountered during postdeposition annealing under Ar-atmosphere of arc deposited TiSiCN coatings, starting at ~800 °C [20].

The low deposition temperatures typically employed with PVD, from room-temperature to ~700 °C make it difficult to produce α-$Al_2O_3$. Instead one of the non-equilibrium phases such as κ or γ is normally obtained [18,21-24]. In addition to low temperatures leading to less diffusion, promoting the metastable phases, small grain sizes also promotes the γ–phase, having lower surface energy than α-$Al_2O_3$ [25]. All the metastable alumina phases transform into corundum either by direct transformation or by different routes involving intermediate phases. The main concern with these transformations is the volume contraction upon phase transformation, which leads to crack networks in the coating [10,26-28]. This makes the non-thermal equilibrium phases less suitable for many metal cutting applications, if not stabilized, due to the high temperatures created in the cutting zone. Upon alloying with small amounts of Si the transition from γ to α is shown to be moved to higher temperatures [29-32].

Different approaches for achieving PVD α-$Al_2O_3$ have been taken throughout the years, such as growth on isomorphic α-$Cr_2O_3$ seed layer [33-36] and alloying with materials stabilizing the corundum phase, i.e. $Cr_2O_3$, as reported for sputtering [37,38] and arc deposition [39-41]. Recent studies also indicates that small Fe-addition can help nucleate the corundum phase [42-45].

By using HIPIMS, semi porous films with platelet-like grains of pure α-$Al_2O_3$ were also manufactured [46]. Higher deposition temperature ~760 °C in combination with constant ion bombardment during reactive magnetron sputtering has also resulted in α-Al$_2$O$_3$ [21]. The type of ion bombardment, metal or Ar$^+$, also need to be considered, in addition to deposition temperature. Higher ion energies of Al-ions, thus penetrating to larger depth and thereby enhancing the diffusion and defect annealing in a larger volume, has been suggested to be the reason for forming α-Al$_2$O$_3$ over γ-Al$_2$O$_3$ in filtered cathodic arc setups [47,48]. This is supported by calculations, where γ-Al$_2$O$_3$ is shown to be more sensitive to ion bombardment than α-Al$_2$O$_3$ [49].

While α-(Al,Cr)$_2$O$_3$ phase has been synthesized from compound cathodes by cathodic arc deposition in industrial equipment, there are still many challenges with respect to process stability and optimized coating properties. The reactive arc deposition from mixed AlCr cathodes in pure oxygen leads to a large amount of metallic droplets in the coating and the formation of oxide islands on the cathode [50,51]. This in turn results in reduced film quality (i.e. porosity and soft intermetallic phases) and uneconomical use of cathode material, respectively, each being detrimental for industrial scale manufacturing. The oxide island problem at the cathode surface exists for a wide range of Al contents in the Al-Cr system, but seems to be reduced for Al-content below 50 at.% [52], possibly related to the shape of the Al-Cr phase diagram causing a reduced amount of Al-melt [53]. The problems with oxide islands may be avoided if the Al-rich melt temperature is below the formation temperature needed for the corundum structure, i.e. 85 at.% Al [51]. However, increasing the Al-content in the cathode leads to more droplets, increased surface roughness and the formation of a metastable B1-like cubic oxide structure [54]. This phase, identified by Khatibi et al. [55], competes with the corundum phase in the (Al,Cr)$_2$O$_3$ system if deposited with PVD, both with magnetron sputtering and with cathodic arc [40,56]. The B1-like phase forms instead of the low temperature γ-Al$_2$O$_3$ phase, which is normally obtained with PVD at ~ 500 °C deposition temperature from elemental Al-targets [21,22,24]. The α-

(Al,Cr)$_2$O$_3$ phase formation in PVD-deposition has become increasingly intriguing by observation of transition from cubic B1-like structure to α-(Al,Cr)$_2$O$_3$ during growth [54,57]. Transformation from γ to α- Al$_2$O$_3$ during growth with filtered cathodic arc has also been reported [58]. Even though predicted for bulk [59,60] it has not been possible to verify spinodal decomposition of the α-(Al,Cr)$_2$O$_3$ in arc deposited coatings [41,54].

If Si is added to the Al$_x$Cr$_{1-x}$ cathode, the arc-process is stabilized, creating a more even erosion of the cathode, allegedly without being incorporated into the coating, possibly due to formation of volatile SiO species [61]. The change in cathode surface has been explained with lowering of the solidification temperature of the melt at the cathode surface, below the formation temperature of the corundum phase. Another study showed lower cathode erosion rates and less oxide island formation with the addition of either 5 at.% Si or B in the AlCr cathode, but these species were incorporated in the coating [62]. A plasma study confirmed the existence of volatile SiO species under DC cathodic arc conditions from an Al$_{70}$Cr$_{25}$Si$_5$ cathode [63]. The more even erosion of the cathode surface, the larger is the industrial advantage for process optimization. However, in the light of all studies, it seems like the Si content in the cathode, as well as the process parameters, need to be balanced carefully. As mentioned above, several studies with Si added to the cathode/target also show Si in the coating, both in depositing (Al$_x$Cr$_{1-x}$)$_2$O$_3$ and Al$_2$O$_3$. The Si in the coating then stabilizes the cubic B1 like phase, γ-phase or a mullite phase, instead of the desired corundum phase [30,31,63,64] It remains to be understood how the cathode Si-content in combination with O$_2$-flow governs the Si-uptake in the coating and its effect on their crystal structure, hardness and E-modulus.

Correspondingly, this work explores the influence of adding Si to different AlCr cathodes and the resulting coating properties. The thin films were deposited by cathodic arc deposition, using the same type of deposition equipment as in ref. [61]. The influence of different oxygen flows is also considered. We show with chemical analysis techniques that Si is incorporated into all studied coatings, independently of the Al/Cr content in the cathode or oxygen flow. The α-$(Al,Cr)_2O_3$ coating, change phase to cubic B1-like-$(Al,Cr)_2O_3$ upon incorporating Si. Change from B1-like to corundum during growth of Si-free $(Al,Cr)_2O_3$ is also observed. The hardness is slightly reduced when alloying Si to the coating.

## II. EXPERIMENTAL DETAILS

The coatings were deposited in an Innova Oerlikon Balzers equipment with cathodes (160 mm diameter) on the upper half of the chamber. The substrates, both Si (001) and WC-Co substrates (10 wt. % Co, ground but not polished surface), were mounted on the upper half of the carousel for two fold rotation at 1.5 rpm. The substrates were cleaned prior to deposition with ethanol and isopropanol. The depositions started with heating to deposition temperature 575 ± 25 °C, followed by Ar plasma etching for 25 min prior to deposition of a TiAlN starting layer of ~250 nm (one $Ti_{50}Al_{50}$ cathode). An oxygen-nitrogen gradient binding layer was then deposited in order to improve oxide adhesion. This was followed by the pure oxide layer deposition during 115 min (2 cathodes/process). The bias voltage applied to the substrate table during the oxide deposition step was symmetric bipolar -120V, 36 µs negative and 4 µs positive (i.e., 90% duty cycle), resulting in a pulsing frequency of 25 kHz. The arc sources were run with pulsed current, max-value 180 A, low current 67% of max value with a duty cycle of 50% giving in average 150 A, at a pulsing frequency of 1 kHz. The distance between cathode and substrate was 14 cm. Base pressure of the system before deposition start was 0.05 Pa and the deposition pressure was between 0.5 and 3 Pa depending on O$_2$ flow. The oxide deposition took place under pure oxygen atmosphere, flow controlled, with 300 (low), 600 (mid), or 1000 (high) sccm of O$_2$ flow. Two coatings were deposited as reference samples of the pure binary Al-Cr-O oxide, having the nominal cathode compositions of Al$_{50}$Cr$_{50}$ and Al$_{70}$Cr$_{30}$. To the Al$_{50}$Cr$_{50}$ base case, 5 and 10 at.% Si was added respectively to the cathode upon reducing the Al and Cr with equal amounts, keeping the Al/(Cr+Al) at 50 at.%. For the Al$_{70}$Cr$_{30}$ case, 5 at.% Si was added by reducing the Cr-content with the same amount, in order to align with previous published work [61]. New cathodes were used for runs using Al$_{47.5}$Cr$_{47.5}$Si$_5$, and Al$_{70}$Cr$_{25}$Si$_5$ compositions with low oxygen flow, and for runs using Al$_{50}$Cr$_{50}$, Al$_{45}$Cr$_{45}$Si$_{10}$, and Al$_{70}$Cr$_{30}$ compositions at mid oxygen flow. All cathode compositions were operated with the reference mid oxygen flow (300 sccm per active source), giving rise to 5 cathode compositional coating variants. The two cathode setups having 5 at.% Si were also run with the two other oxygen flows, starting from the lowest oxygen flow in order to minimize cathode contamination after successive depositions. Between each run, the cathodes were also sand blasted (with Al$_2$O$_3$) in order to remove oxide island formed on the cathode surface. In total, this sum into 5 compositional variants and 3 oxygen flow variants. 70 at.% Al and 5 at. % Si and low oxygen flow was tested but the arcing was not stable and the process was aborted, thus not further reported. The coating variants will in this work be denoted with the nominal cathode content of Al and Si in at.% as well as low, high or mid O$_2$ flow.

The crystallographic structure of the coatings was evaluated using X-ray diffraction (XRD) and transmission electron microscopy (TEM). Grazing incident XRD was done in a Panalytical Empyrian machine with a 1.5° incoming angle, using line focus with a parabolic multilayer mirror, 1/4 ° divergence slit and a 2 mm masking on the incoming beam and a parallel plate collimator (0.27° acceptance angle) followed by a 0.02 mm Nickel foil on the diffracting side with an PIXcel$^{3D}$ detector in open mode. The step size in 2θ was set to 0.07°/step with a count time of 15 s/step, in a 18-80 2θ range. Peak positions were determined by peak fitting in Origin on the rawdata containing intensity from both Cu K$_{\alpha 1,2}$. All XRD measurements were performed on coatings deposited on WC substrates unless other stated.

The TEM imaging and selected area electron diffraction (SAED) pattern were done in a FEI Tecnai G2 TF20 UT instrument equipped with a field emission gun at operating voltage of 200 kV. High resolution EDX mapping and was done in the double-corrected Linköping FEI Titan$^3$ with field emission gun at operation voltage of 300 kV with super X EDX-detector, drastically reducing the measuring time. The TEM-micrographs are oriented with up being the growth direction.

The chemical composition of the coatings was evaluated with energy-dispersive X-ray spectroscopy (EDX) and elastic recoil detection analyses (ERDA). The EDX was done with an 80 mm$^2$ X-Max Oxford Instrument equipment installed on a Leo 1550 Gemini field emission gun scanning electron microscopy (FEG-SEM) with 10 kV acceleration voltage and working distance of 8.5 mm. The measurements were performed on the coatings deposited on hard metal (top view) in order to be certain that the measured Si-content was not from the substrate. The measurement area was 1 mm$^2$. The O-content in the coatings was compared against an α-Al$_2$O$_3$ (0001) reference sample measured at the same time.

ERDA was performed with 36 MeV $^{127}$I$^{8+}$ iodine ions in a tandem accelerator (Uppsala, Sweden). The detector was a combined TOF-E spectrometer. The recoil detection angle between incident beam and detector telescope was 45º with an incidence angle of 67.5° with respect to the sample surface normal. The acquisition depth was ~300 nm and the measurements were performed on post-deposition polished coatings surfaces deposited on WC-Co as well as on Si-substrates. The measurement accuracy for oxygen was ~5%, the precision ~1.5% and the detection limit of a material was ~0.1%. The acquired data was converted to depth profiles using the software Conversion of Time-Energy Spectra (CONTES). Further description of the used analysis technique and facility can be found elsewhere [65,66].

The bonding states and the chemical composition of the materials in the coatings were evaluated with XPS analyses in an Axis Ultra DLD instrument from Kratos Analytical using monochromatic Al Kα radiation (hv = 1486.6 eV), with the base pressure during spectra acquisition of $1.1 \times 10^{-9}$ Torr ($1.5 \times 10^{-7}$ Pa). The anode power was set to 150 W. Surfaces were analyzed both in the as received state as well as after the $Ar^+$ sputter etching performed with 4 keV ion energy and 12.7 mA/cm$^2$ ion density in three 15 min long intervals to ensure that the steady state was reached on these demanding high surface roughness film. After that, a final sputter step consisting of 5 min etch with 500 eV $Ar^+$ was applied in order to minimize the sputter damage [67,68]. The ion incidence angle with respect to the sample normal was 70°, and the beam was rastered over a 3×3 mm$^2$ area, corresponding to the removal of 160 Å/min as determined by cross-sectional SEM fracture cross-section on the reference $Ta_2O_5$ samples. The area analyzed by XPS was 0.3×0.7 mm$^2$ and centered in the middle of the ion-etched crater. The analyzer pass energy was set to 20 eV, which results in a full width at half maximum of 0.55 eV for the Ag $3d_{5/2}$ peak. A floodgun was used to compensate for positive charge build up during measurements. Due to the inhomogeneous sample morphology with metallic droplets embedded into an insulating oxide film calibration of the binding energy scale was performed in a non-standard way. All spectra were aligned to the Al 2p contribution due to Al oxide positioned at 74.4 eV. This approach was chosen since no C 1s peak from adventitious C was present after the sputtering step and recent work showed its unreliability as a calibration source, especially for non-conducting samples [69,70]. Calibration to Ar 2p peak was attempted but it resulted in too high BE values for the Al 2p oxide peak. It has been shown that the Ar 2p binding energy depends on the environment Ar atom is implanted in [71]. Our calibration procedure results in Cr 2p and O 1s BE values for oxides which are consistent with the NIST database. Quantitative analysis was performed using the Casa XPS software and sensitivity factors supplied by the manufacturer.

Since the surface roughness was larger on the WC-Co-substrate than the Si wafers, most of the chemical measurements were also repeated on coatings deposited on Si-substrate. The coatings were also fine polished manually, top down, with 1μm diamond (3M Film) paper in order to remove the rough, as deposited, surface (i.e. droplets). The surface roughness consideration was more important for ERDA and XPS-measurements since these techniques have smaller acquisition depth than EDX. In the later case, the probing depth of the electron beam was within the oxide coating since no Ti-signal was observed (TiAlN-starting layer).

The loading's hardness was measured with a UMIS2000 nanoindentation system equipped with a Berkovich indenter. 32 quasi static (closed loop) indents were made on tapered polished surfaces of the as-deposited coatings with a maximum load of 10 mN resulting in indentation depth well below 10 % of the coating thickness (max depth ~137 nm for the softest sample). The data were analyzed by the technique of Oliver and Pharr [72] and reported here are the mean value and the standard deviation from the at least 23 indents. Young's modulus was calculated based on the mechanical properties of diamond and an assumed constant poission ratio of 0.17 for the coatings. The machine compliance was calibrated to 0.2 nm/mN.

## III. RESULTS

### *A. Phase analysis*

Figure 1 shows grazing incidence XRD diffractograms for the as-deposited coatings. Figure 1 a) and b) show the diffractograms for coatings with 45-50 at.% nominal Al-content in the cathode, and Figure 1 c) and d) show diffractograms of coatings with higher Al-content, 70 at.%. If not stated otherwise, the nominal compositions are used in this paper. Thus Figure 1 a) and c) diffractograms show the effect of adding Si in the cathode on the coatings, whereas Figure 1 b) and d) show the effect of varying the oxygen flow when having 5 at.% nominal Si content in the cathode. The positions of the WC-substrate peaks are marked with black arrows at the bottom of each graph.

Figure 1 a) shows that the coating deposited with nominal $Al_{50}Cr_{50}$ cathode composition and 600 sccm flow (mid $O_2$) (bottom green line) have diffraction peaks matching a corundum (α) solid solution of ~50/50 $Al_2O_3$ and $Cr_2O_3$. The peak positions are just in between those of the end constituents PDF-cards (00-046-1212 and 00-038-1479, respectively) following Vegard's law. All such peaks are marked with boxes having sides positioned at the 2θ-angles of the pure phases, with a height ~ corresponding to the intensity of the PDF-card. In addition there are substrate, TiAlN and cubic $(Al,Cr)_2O_3$, peaks. The intensity of what could be attributed to the $20\bar{2}2$ peak at 45° for the solid solution corundum structure is unexpectedly high for this (according to the PDF-card) low intensity peak. However, this peak also matches the 200 peak of a cubic oxide solid solution phase of the same composition as the corundum structure, as explained in the work by Khatibi et al. [55]. This would result in a unit cell of 4.024 Å (SAED in TEM give 4.08 Å). The corresponding 220 peak for the cubic oxide phase is masked behind higher intensity corundum peak at 65.5°. In addition to the oxide phase the $Ti_{0.5}Al_{0.5}N$ starting layer is visible with 200 diffraction peak at 43.4°. Three of the main substrate WC-reflections are detected in this diffractogram at 31.5°, 35.7°, and 48.3° (reference position from XRD-measurements on bare WC-substrate). The coatings deposited from $Al_{47.5}Cr_{47.5}Si_5$ mid $O_2$ (orange line) and $Al_{45}Cr_{45}Si_{10}$ mid $O_2$ (blue line) cathodes both show a changed crystallographic structure compared to the coating deposited with pure $Al_{50}Cr_{50}$ cathodes. The solid solution α-$(Al,Cr)_2O_3$ peaks have disappeared and the ordered metal vacancy-stabilized cubic B1-like structure, previously reported of by Khatibi et al. [55], [40] and Alling et al. [56] is visible. For the coating deposited with 5 at.% Si in the cathode (orange line), the broad peak at ~44° is a convolution of the 200 $Ti_{0.5}Al_{0.5}N$ peak 42.9°, a small contribution from the oxide or an intermetallic phase (44.2°), and the 200 peak (45°) for the cubic B1-like solid solution oxide phase. The convolution is not clearly visible for the higher order diffractions and the peak at 55.6° matches best (0.1% deviation) the 220 peak of the cubic B1-like oxide structure. The shoulder at 36.6° could correspond to the 111 peak for $Ti_{0.5}Al_{0.5}N$, but would need to include other diffractions as well, e.g., WC-substrate and 111 of the cubic oxide phase expected to be found ~38.6°, in order to explain the width of the high intensity shoulder. The coating deposited with 10 at.% Si in the cathode (blue line) shows similar diffraction peaks as for the 5 at.% Si coating but with more distinct peaks. Some additional peaks were observed, which could not be explained by just the nitride starting layer followed by an oxide layer. Due to the extra unidentified peaks, some having possible overlap with the WC-substrate peaks (especially for the 48.3° peak), an XRD-measurement was done on the 10 at.% Si-coating deposited on Si(001) substrate (red line). This diffractogram confirms the existence of an intermetallic phase, cubic-$Cr_3Si$ (PDF- 00-007-0186), which was (previously) difficult to detect on WC-substrate.

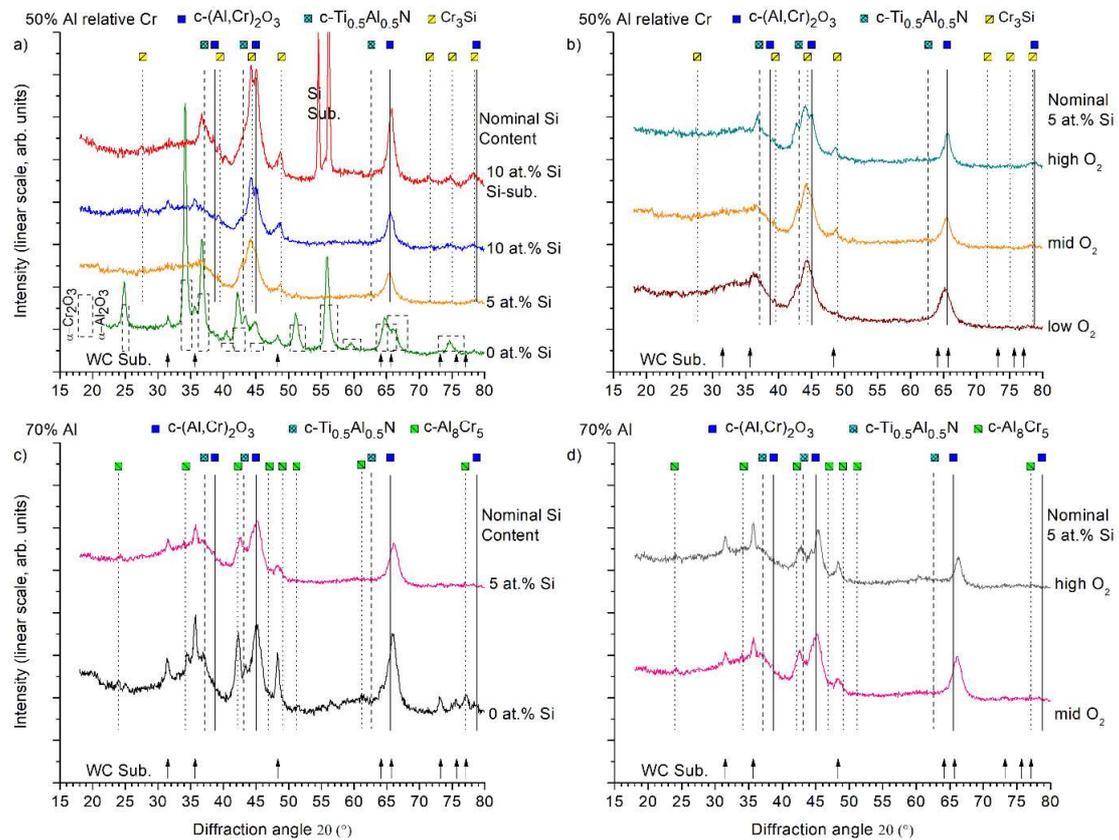

**Figure 1.** Grazing incidence XRD for all evaluated coatings. 1.5° incident beam angle, WC/Co-Substrate for all except one coating (red line). a) influence of Si-content at 50 at.% Al relative Cr content at fixed mid $O_2$ flow, b) influence of $O_2$ flow at fixed 5 at.% Si for 50% Al relative Cr content. c) influence of Si-content at 70 at.% Al metal with fixed mid $O_2$ flow, d) influence of $O_2$ flow at fixed 5 at.% Si for 70 at.% Al metal. Curves shifted for clarity. For color reference, we refer to the online-version.

Thus, the peaks at ~48.5° originate also from the coating, besides WC-substrate, since they are present also for the coating deposited on Si-substrate. In the 10 at.% Si coating on Si-substrate (red line) the peak is found at 48.7°, the same value as for that peak when the coating was deposited on WC-substrate (blue line). This corresponds well (-0.4%) with the 211 diffraction peak at 48.902° (50% normalized intensity) for the Cr₃Si-phase according to its PDF-card. The main diffraction peak for the Cr₃Si phase according to PDF-card is the 210 peak at 44.393°. This peak is clearly visible at 44.3° for the two coatings deposited with 10 at.% Si in the cathode on WC and Si substrate. This phase may also explain the high peak observed at slightly lower angle, 44.2°, for the coating with 5 at.% Si in the cathode.

Figure 1 b) shows grazing incidence XRD diffractograms for the coatings deposited with nominal $Al_{47.5}Cr_{47.5}Si_5$ cathode composition with different oxygen flows, low, mid, and high, from bottom to top. All the three coatings show diffraction peaks from the cubic defect-stabilized B1-like solid solution of $(Al,Cr)_2O_3$ with equal amount of Al and Cr, marked with solid line and blue square in the figure. Except for the absence of a distinct 111 peak, the two primary peaks related to that phase become sharper and better aligned, to the angles expected for stoichiometric 50/50 compound, with higher $O_2$ flows. Since the two highest intensity WC-peak (at 31.5° and 35.7°) are not clearly visible, the peak at 48.6°-46.8° is attributed to the 211 diffraction of $C_3rSi$-phase. The $Cr_3Si$ main reflection 210 results in the highest intensity peak, positioned from 44.3° to 44.1° with increasing $O_2$ flow.

Figure 1 c) shows grazing incidence XRD diffractograms for the coatings deposited with nominal $Al_{70}Cr_{30}$ and $Al_{70}Cr_{25}Si_5$ cathode composition, respectively, both with mid $O_2$ flow, from bottom to top. Increasing the Al content in the cathode leads to that even the pure $(Al,Cr)_2O_3$ coating forms predominantly cubic defect stabilized B1-like solid solution of $(Al,Cr)_2O_3$, but shifted to larger angles 45.25° and 65.83°, due to the higher Al content. This results in a unit cell of 4.00-4.01 Å. The solid line and blue square still mark the $(Al,Cr)_2O_3$ with Al and Cr in equal amounts. There is also an indication of a small fraction of α-$(Al,Cr)_2O_3$ phase, as the four main

diffraction peaks, 012, 104, 110, and 116 at 25.04°, 34.52°, 37,02°, and 56.24°, respectively are detected. An intermetallic cubic $Al_8Cr_5$ phase (pdf-card 47-1466) is consistent with its main peak at 42.31° as well as the minor diffractions at 24.97° and 51.41°. WC-substrate peaks are more visible for these coatings compared to the Cr-richer ones, influenced by the lower mass absorption coefficient of Al compared to Cr, resulting in larger X-ray penetration depth for Al-rich coatings. Upon adding Si to the coating, the last traces of corundum structure disappear and the cubic B1-like oxide phase is the dominant phase in the coating.

Figure 1 d) shows grazing incidence XRD diffractograms for the coatings deposited with nominal $Al_{70}Cr_{25}Si_5$ cathode composition, with mid and high $O_2$ flow, bottom and top respectively. The cubic defect stabilized B1-like solid solution of $(Al,Cr)_2O_3$ is attributable to the peaks at 45.1° and 66.04° for the mid $O_2$ flow coating and 45.45° and 66.32° for the high flow coating. The coating made with high $O_2$ flow show more distinct oxide peaks and lower FWHM.

Figure 2 a) shows dark field TEM and SAED of a pure $Al_{50}Cr_{50}$ coating grown with 600 sccm (mid) oxygen (green diffractogram in Figure 1 a), where it is visible that the oxide starts to grow in a cubic B1-like structure and that larger corundum-structure grains form after ~0.6 µm of oxide film growth. The cubic oxide is slightly textured in the 200 out-of-plane direction. The calculated unit cell for the oxide phase based on the 200 and 220 diffraction is 4.08 Å. The existence of the cubic oxide phase also in this coating explains the relatively high XRD peak at 45° in Figure 1a (green line).

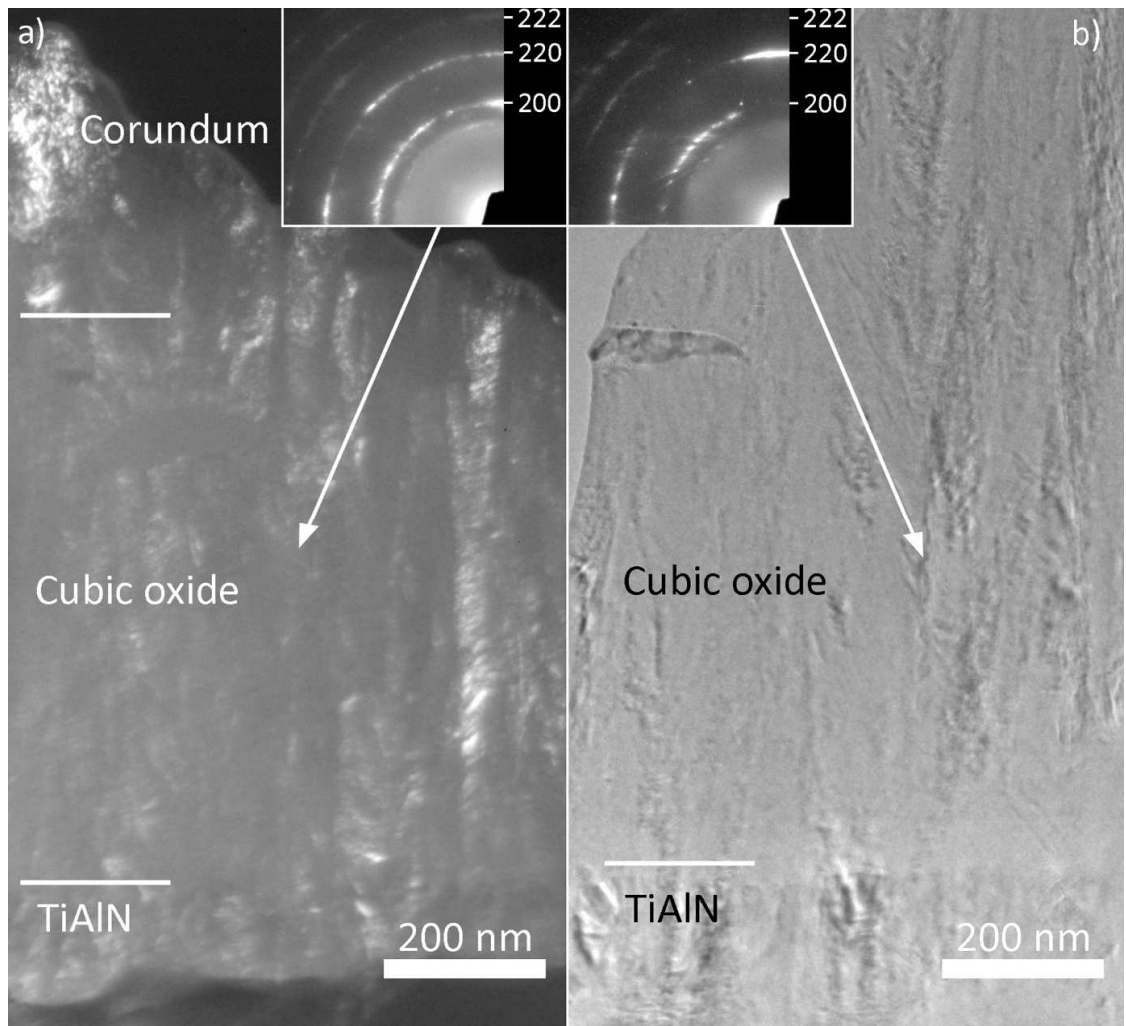

**Figure 2.** a) TEM micrographs (dark field) of (Al0.5Cr0.5)2O3 with SAED of cubic oxide phase with top part of the coating being in corundum phase (SAED not shown). b) TEM-micrograph of (Al0.475Cr0.475,Si0.05)2O3 high O2 flow with corresponding SAED showing the oxide coating with 220 texture in the growth direction.

Figure 2 b) shows TEM and SAED for the coating deposited with 5 at.% added Si in the cathode and high $O_2$ flow, turquoise diffractogram in Figure 1 b). The B1-like cubic oxide phase extends now throughout the entire coating thickness is both samples. The texture changed upon adding Si, to predominantly 220 along the growth direction. Neither of the two cubic oxides show any clear 111 diffraction ring, but rather an increased continues background starting at slightly smaller radius in the SAED than where the 111 should appear in order to match the 200 and 220 diffractions. This is also in accordance with XRD, where a broad intensity increase is seen below ~38°. The 111 indexing is therefore omitted in the presented SAED. SAED from the TiAlN starting layer (not shown) show contrary to the cubic oxide coating a clear 111 diffraction matching the unit cell obtained from 200 diffraction of the TiAlN phase.

Figure 3 a), shows TEM and SAED of the coating made with $Al_{70}Cr_{30}$ cathode composition, mid-$O_2$ flow (black diffractogram in Figure 1c). The oxide coating is found to be predominantly B1-like cubic $(Al,Cr)_2O_3$ as shown by the SAED from a droplet-free region of the oxide coating. No strong texture is observed for the 200 and 220 rings. The possible 111 diffraction ring remains diffuse and of low intensity. The calculated unit cell, based on the 200 and 220 diffraction rings, is 3.99 Å. For the TiAlN start layer, the unit cell is 4.14 Å based on (not shown) SAED 200 diffraction (exhibiting the highest intensity), in TEM. This correspond well to the small peak at 43.43° (4.16 Å) in the corresponding XRD diffractogram. Figure 3 b) shows TEM and SAED of the coating made with $Al_{70}Cr_{25}Si_5$ cathode composition, mid $O_2$ flow (pink diffractogram in Figure 1 c). The SAED from the top part of the oxide coating shows clear 200 and 220 rings. The calculated oxide unit cell, 4.02-4.00 Å for XRD and 4.00-3.99 Å from SAED in TEM, show good agreement.

The SAED background is higher again, below typical 111 ring radius, which corresponds to the increase in intensity seen below ~38° in XRD-diffractograms for cubic oxide containing films (Figure 1). SAED on a Si-rich droplet, showing many small grains, Figure 3 b inset (top right), gives rise to several individual diffraction dots with the main part in the 200 d-spacing range of 2.14-1.88 Å, corresponding to a 2θ range in XRD of 42.2-48.3° (with the maximum intensity peak being at 42.5°). This helps to explain the many and broad XRD-peaks in this range, in addition to the known oxide phase and TiAlN-start layer, also giving diffraction at distinct position in this range (the dotted half circles in the SAED serves as a guide for the eye, matching the 200 and 220, respectively, of the cubic oxide phase). In addition to the Si-rich elongated droplet, often small and dark in the overview TEM micrograph, there are also larger flat white droplets as well as round droplets in this coating.

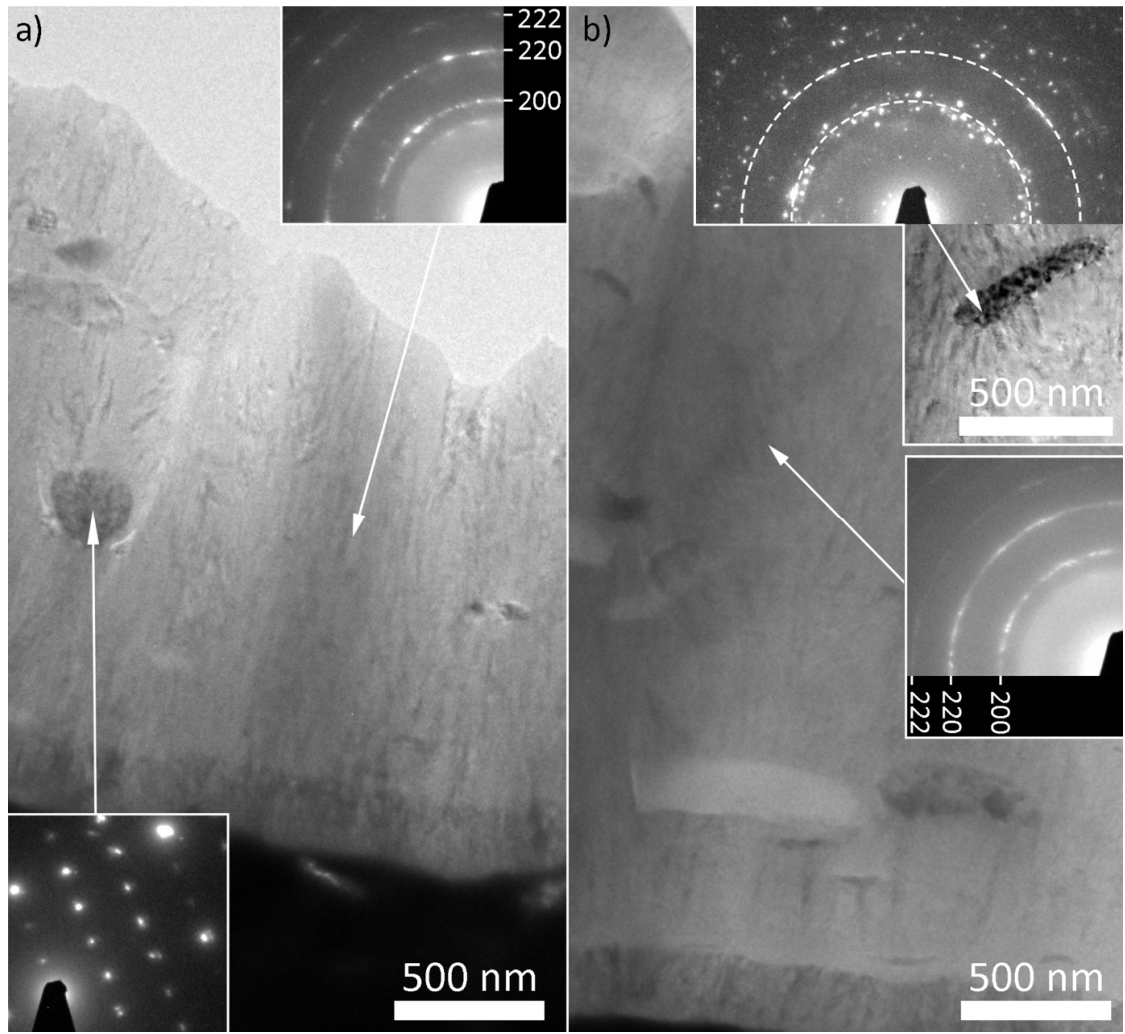

Figure 3. a) TEM micrographs of cubic-(Al0.7Cr0.3)2O3 with SAED of cubic oxide phase at the top and SAED of highly crystalline Cr-rich droplet bottom. b) TEM-micrograph of cubic-(Al0.7Cr0.25Si0.05)2O3 mid O2 flow. Bottom SAED of the cubic oxide. Top inset is from the same coating but different location, with SAED of nanocrystalline Si-rich droplet (dotted circles marks 220 and 200 for the cubic oxide for comparison).

In order to better understand the Si distribution and the different droplet-types, additional high-resolution STEM and EDX mapping was performed on the coating deposited from $Al_{70}Cr_{25}Si_5$ cathode composition, and mid $O_2$ flow. Figure 4 a) shows an example of a round, ~220 nm in diameter, Cr- and Al-rich droplet with small amounts of Si. The smaller flat droplet down to the left is rich in Si with ~24 at.%, but due to the thickness of the sample at this position this value is probably underestimated due to overlap with the oxide coating which has a lower Si content. A scan down to the right in the image of the oxide coating gives a 54 at.% O, 1.5 at.% Si, 33 at.% Al, and 12 at.% Cr, showing that the metal fraction of the oxide aligns well with cathode composition. Figure 4 b) shows another type of droplet, similar to the one discussed in conjunction with Figure 3 b), with two types of small grains embedded in a larger droplet. The light-contrast ones are rich in Si and Cr, 16 and 53 at.% respectively, giving an approximate composition matching the $SiCr_3$-phase previously discussed. The dark areas on the other hand are rich on Al with 34 at.% and have Cr and Si in lower content, 24 and 10 at.% respectively, the balance being oxygen and trace elements of resputtered Ti and W from TEM-grid and substrate, respectively. The mapping confirms the large difference in composition that can be observed between the main coating and the droplets and shows Si in all parts of the coating to varying degree. No variation of the Si content in the oxide coating could be detected below the ~5 nm scale.

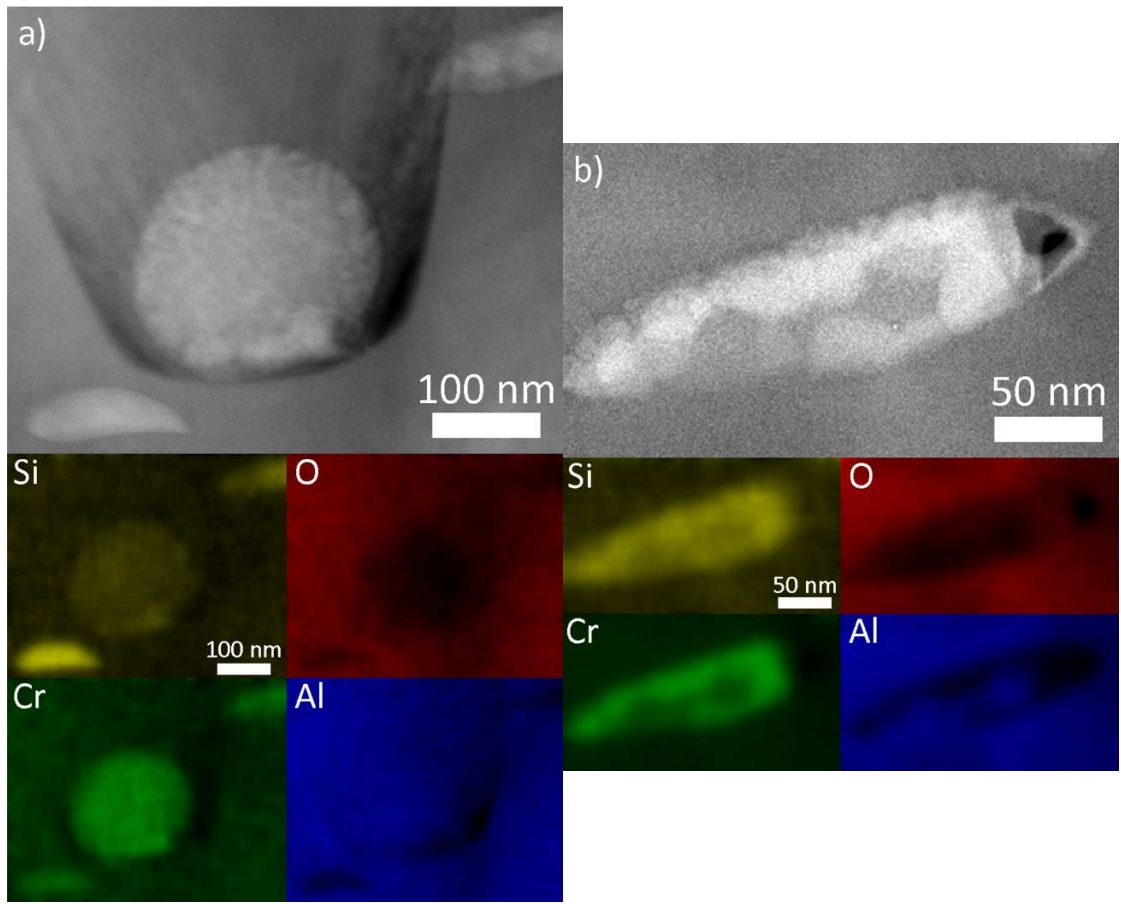

Figure 4. HRSTEM HAADF micrographs with corresponding EDX-maps of Si and Cr-rich droplets from cubic-$(Al_{0.7}Cr_{0.25}Si_{0.05})_2O_3$ mid $O_2$ flow coating. a) Large round Cr and Al-rich as well as flat Cr and Si-rich droplets and b) nanocrystalline Si and Cr-rich droplets. For color figure, we refer to the online version.

## B. *Elemental analysis*

The samples were characterized for chemical composition with EDX and ERDA. EDX was done on the samples on hard-metal substrate in order to eliminate the risk of Si-signal from the substrate. Table 1 shows the composition acquired with EDX. In addition to the measured elemental content the tables provides additional calculated atomic concentrations % of Al/(Cr+Al), metal fraction % of Si and metal to oxygen fraction. Five coatings were deposited with Al/(Cr+Al) content of 50% in the cathode. Three coatings had 70% Al in the cathode, but different Cr and Si-content.

The EDX measurements of the Al-content for all these coatings was in close agreement with the respective nominal cathode composition, see Table 1.

Table 1. The chemical compositions of the coatings measured with EDX on WC-substrates. C and N content not stated since ≥0.2 at. % with ERDA.

| Coating variants | Elemental composition [at. %] | | | | $\dfrac{Al \cdot 100}{Al + Cr}$ | $\dfrac{Si \cdot 100}{Cr + Al + Si}$ | $\dfrac{(Al + Cr + Si)}{O}$ |
|---|---|---|---|---|---|---|---|
| | Al | Cr | Si | O | | | |
| $Al_{50}Cr_{50}$, mid $O_2$ | 20.5 | 20.8 | 0 | 58.8 | 49.6 | 0 | 0.70 |
| $Al_{47.5}Cr_{47.5}Si_5$, low $O_2$ | 19.1 | 19.9 | 1.4 | 59.6 | 48.9 | 3.4 | 0.68 |
| $Al_{47.5}Cr_{47.5}Si_5$, mid $O_2$ | 20.4 | 19.8 | 1.4 | 58.4 | 50.8 | 3.3 | 0.71 |
| $Al_{47.5}Cr_{47.5}Si_5$, high $O_2$ | 19.9 | 20.9 | 1.4 | 57.8 | 48.7 | 3.3 | 0.73 |
| $Al_{45}Cr_{45}Si_{10}$, mid $O_2$ | 19.8 | 19.5 | 3.2 | 57.6 | 50.5 | 7.4 | 0.74 |
| $Al_{70}Cr_{30}$, mid $O_2$ | 28.9 | 12.0 | 0 | 59.2 | 70.7 | 0 | 0.69 |
| $Al_{70}Cr_{25}Si_5$, mid $O_2$ | 27.9 | 10.2 | 1.8 | 60.2 | 73.3 | 4.5 | 0.66 |
| $Al_{70}Cr_{25}Si_5$, high $O_2$ | 28.4 | 10.1 | 1.4 | 60.1 | 73.8 | 3.4 | 0.66 |

Similarly the measured O-content is close to the expected value for corundum stoichiometry 60 at.% (for $Me_2O_3$). The 50%-Al coatings show a slightly larger under stoichiometric deviation than the High Al-containing coatings. The maximum difference in O content between the coatings and the expected stoichiometric value is 2.4 at.%, which corresponds to a relative deviation of 4 % to the expected value. Knowing the measurement uncertainties for light elements with EDX, no further conclusion or trends can be drawn from the O levels with EDX. The EDX show, however, presence of Si in all coatings made from cathodes containing Si, between 3.3 and 7.4 at.% metal fraction Si when adding 5 or 10 at.% in the cathodes respectively. The measured Si content in the coatings is thus reduced somewhat from the nominal Si cathode composition.

The oxygen content measured with ERDA shows a similar trend between the samples as observed with EDX, but the absolute values where generally shifted to lower values, 56.3-57.6 at.% O. The C and N impurities were measured to be on a low level ≥0.2 at.%. In the used setup, Si cannot be accurately resolved from Al for low content like this. The measured composition from different techniques will be further treated in the discussion. The composition measured with ERDA can be found in supplementary material, Table 1.

## C.  Bonding analysis XPS

Figure 5) shows the XPS core level spectra of the Si 2p, Cr2p Al 2p, and O1s. The calibrated spectra are normalized to the intensity of the background on the lowest binding energy and shifted vertically for clarity of view. The left column, Figure a)-d), show the spectra from the samples with 50 at.% Al/(Al+Cr) with varying Si content from 0-10 at.%, green, orange and blue spectra respectively. The right column, Figure e)-h), show the corresponding spectra for the 70 at.% Al coating, with 0 or 5 at.% Si in the cathode, black and pink spectra, respectively. If not specified the referenced binding energies comes from NIST XPS online database [73].

The Si 2p spectra Figure 5 a) show two clear peaks, the higher BE peak due to Si-$O_x$, and the lower corresponds to metallic Si. The oxide peak for the 50 at.% Al is situated at 102.3 eV and does not shift with further addition of Si, but the signal intensity increases significantly. Si 2p signal from $SiO_2$ is tabulated to 103.4-103.8 eV range while SiO is normally found around 102 eV, which indicates that this $SiO_x$ is in between the two phases.

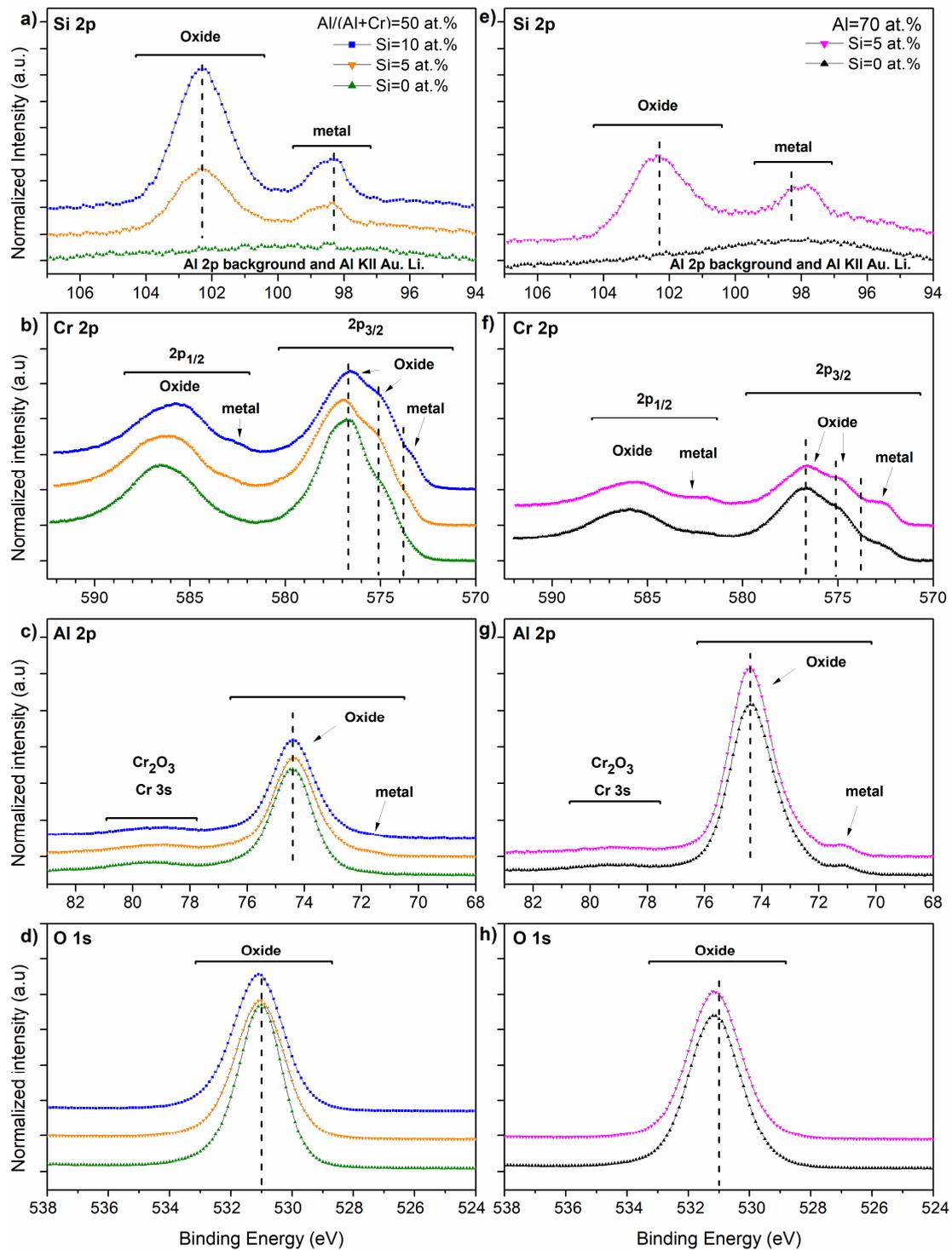

**Figure 5.** XPS spectrum for Si2p (top), Cr2p and Al2p and O 1s (bottom) for different Si-content in the coatings. Left column shows the results for the coatings with 50 at. % Al/(Al+Cr) and the right column shows the results for the coatings with 70 at. % Al in total metal fraction. The results are normalized to the values of low energy background of respective Cr 2p spectrum and shifted vertically to improve visibility.

The Si-Me peak maxima is situated between 98-99 eV (dashed line at 98.3 eV) and includes at least two different contributions, that cannot be explained by the $2p_{3/2}$-$2p_{1/2}$ spin-splitting since the latter requires 2:1 area ratio between two peaks, which is not the case. Instead, we assign this peak to two contributions from Si, found both in Al- and Cr-rich droplets, having different crystal structures. Si-Si is tabulated around 99.3 eV, but is found with values as low as 98.4 eV and other intermetallic compounds are typically found around 99 eV, making this metal peak being at the lower end of tabulated BE values. This is an indirect consequence of our calibration due to the larger than usual observed split between oxide and metal peaks for all our spectra with metal contribution, which leads to lower BE for metallic peaks than normally tabulated. More detailed discussion is provided in section IV.C. Si 2p spectra Figure 5 e), for the coating with 70 at.% Al and 5 at.% Si in the cathode, show an oxide peak which remain at 102.3 eV, but the metallic part shifts to even lower BE, around 97.5-98.5 eV. For both samples without Si in the cathode, there is still a broad peak, more visible for the high Al-containing coating. This background peak is attributed to a combination of Al 2p background and Al KLL Auger line, which becomes more visible with more Al in the coating.

Figure 5 b) show the Cr 2p peak spectra which show several features, owing a high flexibility in Cr oxidation states ($Cr^{+2}$, $Cr^{+3}$, $Cr^{+4}$, $Cr^{+6}$). Due to the complex spectrum of the Cr 2p, with multiple slightly overlapping peaks, the focus in interpretation will be on the higher intensity Cr $2p_{3/2}$. The Cr $2p_{3/2}$ for 50 at.% Al/(Al+Cr) with increasing Si content show 2-3 peaks or shoulders where the main oxide peak is at ~576.7 eV, another oxide peak is found at ~575.1 eV, while a metallic Cr is detected at ~ 573.8 eV (positions marked with dashed lines as guide for the eye).

The expected BE values for a $Cr_2O_3$, $CrO_2$, and Cr are 576.1-576.8 eV, 575.4 eV, and 573.8-574 eV, respectively, which is in the range of the observed values. The Si-free coating shows the sharpest peaks, situated at the highest average BE, with no clear low energy metal shoulder. With increasing Si content, the intensity of the main peak is reduced and slightly broadened, and shifts to slightly to lower BE for measurements done on the 10 at.% Si coating. The Cr $2p_{1/2}$ has similar, but fewer, features as the $2p_{3/2}$ due to increased broadening and lower intensities. The metallic Cr$2p_{1/2}$ peak is clearly visible at ~582.8 eV for the coating with 10 at.% Si in the cathode. Figure 5 f) shows the corresponding Cr 2p spectra for 70 at.% Al coatings. The main oxide peak is situated at 576.7 eV for the coating without Si and shifts by 0.1 eV to lower BE with adding Si. The second oxide peak is significantly more pronounced compared to the 50 at.% Al/(Al+Cr) coatings, but situated approximately at the same BE. The biggest change compared to the 50 at.%-Al coatings is the more visible metallic shoulder on the Cr $2p_{3/2}$ peak which also is shifted to lower BE, ~573 eV. The Cr $2p_{1/2}$ show similar features as for the corresponding 50 at.%-Al coatings, with a broad oxide peak, although now also with clear metallic binding contribution for both coatings.

Figure 5 c) and g) show the Al 2p core level spectra for the 50 at. Al/(Al+Cr) and 70 at.% Al coatings, respectively. The main Al oxide peak is comparable for all coatings and have been shifted to 74.4 eV as part of the calibration. This peak is normally reported in the range of 74 eV- 74.7 eV. The main differences between the spectra is the intensity of the metallic contribution which is detected at ~71.7 eV and ~71.1 eV BE for the 50 and 70 at.% Al coatings, respectively. Both values are well below the reference Al 2p BE, typically in the range of 72.5-72.9 eV, which is a direct consequence of the applied charge correction, where we have intentionally chosen the signal from the dominant chemical species (Al oxide) to serve as a BE reference. The resulting BE of all peaks originating from metal inclusions is thus off by the value set by the difference between metal and oxide charging states during XPS measurement. The broad shoulder at 79.4 eV is attributed to the Cr 3s peak of $Cr_2O_3$, normally situated at 75.1 eV-75.4 eV, yet reported as high as 78.9 eV. The observed shift is attributed to the oxide-metal offset, which is discussed in section IV.C.

Figure 5 d) shows the O 1s peak that is situated at 531 eV and shifted only with 0.1 eV to higher BE for the coating with 10 at.% Si in the cathode compared to the 50 at.% Al without Si. There is also a broadening of the peaks with increasing Si content, probably related to the addition of Si to form additional O bonding. The O 1s bond in $SiO_2$ has a higher binding energy range 532.5-533.2 eV in comparison to O 1s in $α-Cr_2O_3$ and $α-Al_2O_3$, 530.1-530.3 eV and 531.0-531.5 eV, respectively. Figure 5 h) shows the O 1s peak for the 70 at.% Al coatings, which is positioned at 531.1 eV also for the Si-free coating. Adding 5 at.% Si only slightly reduces the intensity of the peak. The $O_2$ flow variants for the 5 at.% Si variants, for both Al-compositions, showed only small changes in addition to the above described trends in this section (spectra not shown).

## D. Hardness measurements

Figure 6) shows the results of nanoindentation hardness measurements on all the coatings. The highest hardness is obtained for the 50 at.%-Al $(Al,Cr)_2O_3$ coating at 30±2.4 GPa. Adding 5 or 10 at.% Si to the 50%-Al coating with mid-$O_2$ flow reduces the hardness to 22.9±1.7 and 23.8±2.4 GPa, respectively. Both the coating with low and high $O_2$ flow and 5 at.% Si have slightly higher hardness than the coating deposited with mid-$O_2$ flow, around 25.5 GPa. For the series with 70 at.% Al in the cathode, the Si-free coating shows the highest hardness at 25.8±2.0 GPa, although lower than the corresponding 50 at.%-Al coating. Adding 5 at.% Si to this coating seems to marginally reduce the hardness, while a minor decrease is observed with high $O_2$ flow, down to 23.1±1.9 GPa. The Young's modulus shows its highest value for the 50 at.% Al and is significantly reduced upon Si incorporation from ~390 to 320 GPa. The 70 at.% Al coating start of on a lower level, 360 GPa, and exhibits a smaller reduction when adding 5 at.% Si in the cathode. The hard-metal substrate was measured in conjunction with the coatings, giving a hardness and Young's modulus of 25±3.4 GPa and 661±59 GPa, respectively.

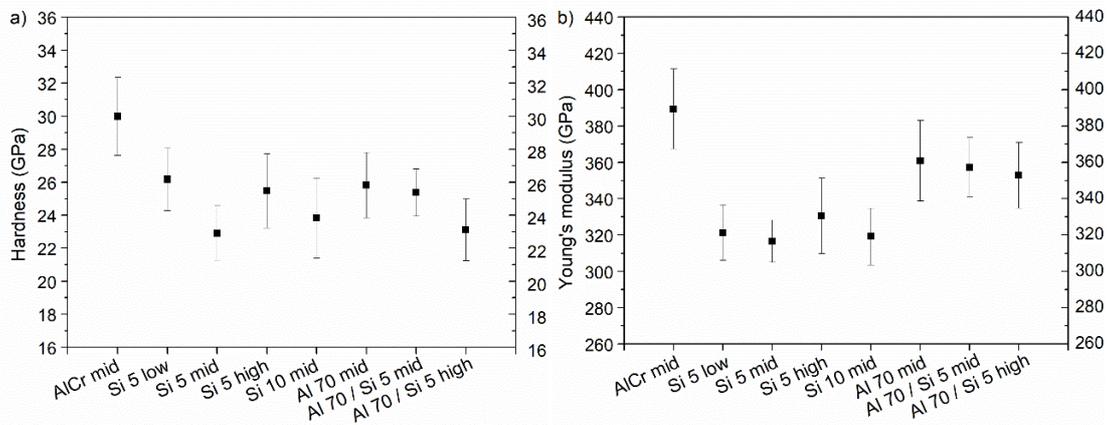

## IV. DISCUSSION

### A. Phase composition

Figure 1 a) and Figure 2 a) show clear formation of the α-$(Al,Cr)_2O_3$ for the coating without Si in the cathode, and to smaller extent the corundum peak is also visible for the 70 at.% Al-containing $(Al,Cr)_2O_3$ coating, in accordance with [39,40]. When Si is added to the cathode, for both low and high Al content, it results in a change of crystallographic structure to B1-like cubic $(Al,Cr)_2O_3$ oxide, Figure 1 a-d) and Figure 2 b) and Figure 3 b). The shift of crystallographic structure to B1-like cubic (Al,Cr)$_2$O$_3$ upon adding 5 at.% Si to ~70 at.% Al-containing AlCrSi cathodes is in line with previous work [61,62]. In pure Al$_2$O$_3$, Si or SiO$_2$ addition lead to the stabilization of another metastable polymorph, primarily γ-Al$_2$O$_3$ [30,74]. In our case, the gamma phase is excluded because of the absent 111 peak at ~19.45° and higher-order peaks, 222 and 333, at corresponding angles. The resulting lattice parameter from XRD (TEM SAED for comparison) for the B1-like cubic oxide with ~50 at.% Al/(Al+Cr) is ~4.02-4.04 Å (4.03-4.09 Å) and for 70 at.% Al in the cathode it is 3.99-4.02 Å (3.98-4.01 Å), respectively. Previous work have reported the unit cell to be ~4.03 Å for ~50 at.% Al and ~4.00 Å for Al-content ~70 at.% [40,55,62,64]; thus, the present work in good agreement with literature, especially for XRD, which is judged to be the most reliable data.

The observed transition from cubic to corundum growth (Al,Cr)$_2$O$_3$ for the 50 at.% Al/(Al+Cr) coating in this work has previously been observed and discussed by Najafi et al. [57]. In the present work, the earlier onset of the corundum structure growth, at ~0.6 µm instead of 2 µm in the work by Najafi et al., may have been promoted by several factors; slightly higher Cr content in the cathode in combination with slightly higher deposition temperature and roughly half the deposition rate. According to other prior work [39,40], higher Cr content and O$_2$ flow should promote the corundum structure formation.

An interesting feature of the B1-like oxide cubic-(Al,Cr)$_2$O$_3$ phase is the absence of a distinct 111 peak. According to the first article proposing this structure for this material system by Khatibi et al. [55], pole figures at the angle corresponding to the 111 diffraction show the expected ring with respect to the (200) plane. This peak is, however, not strongly visible in the present work, neither in XRD nor TEM, as well as in other published work [40,54,75]. Najafi et al. showed drastic reduction of the 111 diffraction upon O-incorporation in $Al_{70}Cr_{30}N$ arc-deposited coatings. Ordering of the cation vacancies on the (111) planes was proposed as an explanation to the low observed intensity of the 111 peak[76]. In works by Koller et al. no 111 peak can be seen for either $Al_{70}Cr_{30}/Al_{75}Cr_{25}$ oxide coating nor when adding 5 at.% Si to the high Al containing AlCr cathode. This is even though the XRD was done on ground powder from the deposited coating, hence loss of intensity due to texture could not be the explanation for the disappearance of the 111 peak [54,62]. In this work a broad shoulder below ~38° could be seen in Figure 1, also seen for lower indices in SAED in Figure 2 and 3, starting around the distance where the 111 ring should appear. The absence of a distinct 111 peak can be due to changes in structure factor or a reduction of long range ordering in the <111> direction. Additional XRD scans with larger 2θ-range on some of the coatings in this work (not shown) shows a possible 222 peak at ~83°. TEM SAED (Figure 2 and 3) also shows this diffraction. This indicates ordering on at least half the distance compared to the (111) planes and/or the sum of the atomic form factors being close to equal for the cation and anion sublattices, thus leading to that all odd diffractions for the sodium chloride B1-structure, 111 and 311, being close to zero in intensity.

The many XRD peaks in the 2θ-range of 42-45.5°, visible in Figure 1, is due to the TiAlN start-layer, the cubic oxide and several intermetallic phases, as most of the plausible intermetallic phases (based on composition) have at least one principal peak in this 2θ range. These intermetallic phases, varying significantly in size and shape, contribute the most to the difficulty to identify the diffraction peaks with certainty, especially shown with the top inset in Figure 3b. The XRD peaks are based on a much larger sample volume than TEM and are therefore a more accurate way of observing the dominant droplet phases under the respective deposition conditions, such as cubic $Cr_3Si$ for Cr- and Si-rich cathodes and cubic $Al_8Cr_5$, with dissolved Si for Al-rich cathodes. Even though the droplet fraction is qualitatively significantly smaller than the oxide coating their XRD peak may still be comparable in intensity due to relatively large-size crystallites (as large as the entire round droplets, i.e. Figure 3 a) compared to the cubic oxide phase, as seen by TEM (Figure 2 and 3).

Thus, Si in the cathode promotes the cubic defect B1-crystal oxide structure and the diffraction peaks become more well defined with increasing $O_2$-flow for both high and mid Al at.% in the cathode. The $O_2$-flow variation does not present a way under these deposition condition to shift the crystal structure to corundum once Si is added to the cathode. The 111 and possibly also the 311 diffraction is suppressed when depositing such coatings, the full explanation for this would merit further work but is outside the scope of this work. The different droplet phases contribute significantly to the 2theta range where the primary cubic oxide peak is found and caution need to be taken when evaluating this range, also considering the importance of wide range XRD scans and local TEM SAED in combination to be able to identify the actual phases.

## *B.   Chemical composition*

The elemental analysis with EDX seen in Table 1 shows compositions that are in line with stoichiometric compounds $Me_2O_3$, within the uncertainty limit of the technique. Si is observed in all coatings where Si-containing cathodes were used and no strong correlation between Si-content and $O_2$-flow could be observed. The small reduction in Si content measured in the coating compared to the nominal cathode composition may be connected to the volatile $SiO_x$ species shown to be present by Zhirkov et al. [63] (in an DC-arc discharge from an $Al_{70}Cr_{25}Si_5$ cathode), although the loss of Si was once again shown to be minor. The local structural and chemical analysis, primarily done with STEM-EDX maps, Figure 4, show that Si is present in the oxide coating on a level comparable to what is measured with larger sample volume techniques, as well as being significantly enriched in droplets. The later observation is consistent with the structural analysis with XRD where Si-containing phases such as $Cr_3Si$ have been detected. The STEM-EDX-map does not show enrichment of Si in grain boundaries, but a rather even distribution in the coating. Segregation of Si to the grain boundaries has been the primary explanation to the location of Si in Si-containing $\gamma$-$Al_2O_3$ coatings, deposited by filtered cathodic arc by Nahif et al. [30]. The grain size of the B1-like cubic $(AlCrSi)_2O_3$ studied in the here presented work is however that small that several grains are superimposed in the TEM-sample. Hence, segregation at the grain boundaries cannot be excluded even though STEM-EDX-maps show no Si fluctuations.

EDX measurements (Table 1) show stoichiometric composition with respect to oxygen. The ERDA results (supplementary material Table 1) show a similar trend, but absolute values that are ~3 at.% lower. The explanation for this difference can be twofold; firstly, ERDA uses a significantly larger measuring surface area than EDX in combination with a ~300 nm measurement depth, giving a larger sample volume. This lead to a higher probability of measuring metallic droplets thus reducing the overall average oxygen content. Secondly, a rigid offset with respect to oxygen due to uncertainty in stopping powers for early transition metal elements in combination with the exactness of the sensitivity factors for O also ad to the stoichiometry uncertainty. The later has resulted in a measured 58 at.% oxygen content on a single crystal sapphire substrate in a similar measurement set-up in another study covering sputtered Al-V-O coatings[77]. Hence, the conclusion is that the measured oxygen content with ERDA is stoichiometric $Me_2O_3$, in accordance with the EDX-measurements, given the measurement uncertainties in combination with a metal fraction from metallic droplets incorporated into the coatings.

## C. Bonding

Al 2p, Cr 2p, and Si 2p spectra typically show presence of both an oxide peak (at higher BE) and a metallic peak at lower BE, Figure 5. The main finding with XPS, Figure 5, a) and e), is the clear oxide and metal component (the latter due to droplets) in the Si 2p core-level peak, where the oxide component is situated at values between the SiO and the $SiO_2$. The metal components for Cr 2p, Al 2p and Si 2p spectra, show an increase in intensity with an increased Al content in the coating, indicating more droplets with increasing Al content, which is in line with our observations by TEM.

After alignment of all spectra to the Al 2p oxide peak, set at 74.4 eV, the other metal oxide peaks are well aligned to the corresponding reference values. However, the metallic peaks are shifted to lower BE than expected, Figure 5). This is explained by the sample morphology, which consists of metallic conducting particles in the form of droplets, dispersed in a dielectric matrix. As droplets are not in a proper electrical contact to the substrate, they also undergo certain degree of charging during XPS measurement, which is different from that in the oxide volume. As a consequence the observed BE difference between metal and oxide peaks differs from the reference values leading to artefacts like the one with too low BE of the metal peaks.

The split between oxide and metal component increases with increasing Al content in the coating, i.e. Figure 5 c) and g). This could either be due to an actual shift in BE, due to compositional change, or a change in charge built up, affecting the measured BE of the sample with increasing Al-content. This kind of non-rigid shift between oxide and metal is also treated in the work of Baer et al. when studying the charge compensation of Au dots on a SiO$_2$ sample [78].

Based on our calibration of the Al 2p peak to 74.4 eV, the corresponding Al 2s peak is situated at 119.3 eV for our corundum-structured sample 50 at.% Al/(Al+Cr). This can be compared with an inter laboratory test performed on α-Al$_2$O$_3$ and two other isolating material [79], where different charge referencing techniques where evaluated. This benchmark resulted in an Al 2s peak at 119.41±0.16 eV, when the samples were corrected toward the binding energy of Au 4f$_{7/2}$, present as surface deposited nanoparticles [79].

The here observed broadening of the Cr 2p3/2 peak, Figure 5 b) and f) can be due to several reasons, Ar+ sputtering damage, compositional variations leading to Cr being present in different oxidations states, or an intrinsic splitting of the Me 2P$_{3/2}$ peaks typically seen for the first row transition metal oxides. The latter has been suggested by Pratt and McIntyre [80] due to a coupling between core hole and an unfilled d-shell. The splitting is also visible in other later works [81,82]. The here studied film does not show the distinct split but a broadening with a lower BE energy shoulder. The broadening of XPS peaks for insulating materials seems to increase with the combined use of grounding and flood gun [78]. Sputter cleaning of the surface also often lead to broadening of the peaks and appearance of new spectral features [68,83,84]. See Figure S2 in supplementary material for the effect of sputtering on the Cr 2p shape for one of the coatings.

### *D.   Coating hardness*

The measured hardness for Si-less 50%-Al oxide in this work is 30±2.4 GPa, which is the maximum hardness values for the studied coatings. With increasing Al content, the hardness values decrease. Adding Si to the two different Al composition led to a reduction in the hardness in most cases. For the sample with 70 at.% Al and high oxygen flow, a ~10 % reduction was observed but a comparable hardness was obtained when depositing with mid oxygen flow, Figure 6 a). The Si-containing coatings with 50 % Al/(Al+Cr) in the coating showed a reduction to similar values as the high-Al content coatings. Another recent work comparing $Al_{70}Cr_{30}$ with and without 5 at.% Si in the cathode show lower absolute hardness and modulus values compared to this work but the same trend, where Si addition lowers the hardness with ~13 % [62]. The reduction in hardness with added Si can be explained by increase in droplet density, a mechanically softer phase. From a coatings mechanical point of view the Si content should therefore be kept as low as possible, while still obtaining the positive effect of less oxide islands at the cathode.

## V. CONCLUSIONS

This work has shown the influence of adding 5 or 10 at.% of Si to the Al-Cr cathode on the $(Al,Cr)_2O_3$ coating properties, Al/(Al+Cr) = 0.5 or Al = 70 at.%, in reactive cathodic arc deposition. Between 3.3 and 7.4 at.% Si metal fraction is incorporated into all studied coatings for a Si content of 5 or 10 at.%, respectively, in the cathode. The incorporated Si content does not change significantly with different oxygen flow, covering a wide range of deposition conditions, from low to high $O_2$ flow. Without Si, $\alpha$-$(Al,Cr)_2O_3$ was obtained for Al/(Al+Cr) = 0.5. Cubic B1-like-$(Al,Cr)_2O_3$ was obtained for coatings including Si. Change from B1-like to corundum during growth of Si-free $(Al,Cr)_2O_3$ was also observed for the Al/(Al+Cr) = 0.5 coating. The hardness is slightly reduced when adding Si in the coating or when increasing the Al-content in the coating. Si is found enriched in both Al-rich and Cr-rich metallic droplets, but can also be found on a lower level (similar to the global average content), without visible segregation on the ~ 5 nm scale, in the actual oxide coating. The positive effect of improved cathode erosion upon Si incorporation has to be balanced against the promotion of the metastable B1-like structure, having lower room temperature hardness, and recently shown inferior temperature stability, compared to deposition of the corundum structure directly.

# ACKNOWLEDGMENTS

The Swedish Research Council (VR, grant number: 621-212-4368) is acknowledged for financial support for L.L's industry PhD studies with AB Sandvik Coromant. We acknowledge the Swedish Government Strategic Research Area in Materials Science on Functional Materials at Linköping University (Faculty Grant SFO-Mat-LiU No. 2009 00971). The Knut and Alice Wallenberg Foundation are acknowledged for supporting the Linköping Ultra Electron Microscopy Laboratory, and a Wallenberg Scholar Grant KAW2016.0358 (L.H.) and an Academy Fellow Grant (P.E.). B.A. acknowledges financial support from the Swedish Research Council (VR) through the International Career Grant No. 330-2014-6336 and by Marie Sklodowska Curie Actions, Cofund, Project INCA 600398, as well as from the Swedish Foundation for Strategic Research (SSF) through the Future Research Leaders 6 program. Support from the Swedish research council VR-RFI (contracts #821-2012-5144 & #2017-00646_9) for the Accelerator based ion-technological center, and from the Swedish Foundation for Strategic Research (SSF, contract RIF14-0053) for the Tandem accelerator Laboratory in Uppsala is gratefully acknowledged. L. H. and G. G. also acknowledge, the Swedish Research Council VR Grant 2018-03957, and the VINNOVA grant 2018-04290.

# Appendix

For supplementary material please refer to the online publication and the therein linked material.

# Supplementary material to:
# Influence of Si doping and $O_2$-flow on arc deposited $(Al,Cr)_2O_3$ coatings

L. Landälv[*,1,2], E. Göthelid[2], J. Jensen[1], G. Greczynski[1], J. Lu[1], M. Ahlgren[2], L. Hultman[1], B. Alling[3], P. Eklund[*,1]

[1] Thin Film Physics Division, Department of Physics, Chemistry, and Biology (IFM), Linköping University, Linköping SE-581 83, Sweden
[2] Sandvik Coromant AB, Stockholm SE-126 80, Sweden
[3] Theoretical Physics, Department of Physics, Chemistry, and Biology (IFM), Linköping University, Linköping SE-581 83, Sweden

**Table S1.** The chemical compositions of the coatings measured with ERDA (on polished Si wafers in order to have as flat surface as possible) Note that the Al+Si is put together since the used ERDA setup couldn't resolve these materials.

| Coating variants | Elemental composition [at %] | | | | | $\frac{Al+Si}{Cr}$ | $\frac{(Al+Cr+Si)}{O}$ | Structure |
|---|---|---|---|---|---|---|---|---|
| | Al+Si | Cr | O | C | N | | | |
| AlCr mid $O_2$ | 20.1 | 22.4 | 57.4 | 0.1 | 0 | 0.90 | 0.74 | Cubic B1→ Corundum |
| AlCr 5% Si low $O_2$ | 21.0 | 21.2 | 57.5 | 0.1 | 0.2 | 0.99 | 0.73 | Cubic B1 |
| AlCr 5% Si mid $O_2$ | 21.6 | 21.8 | 56.4 | 0.2 | 0 | 0.99 | 0.77 | Cubic B1 |
| AlCr 5% Si high $O_2$ | Not analyzed with this tech. On Si substrate | | | | | | | Cubic B1 |
| AlCr 10% Si mid $O_2$ | 22.7 | 20.8 | 56.3 | 0.2 | 0 | 1.09 | 0.77 | Cubic B1 |
| $Al_{70}Cr_{30}$ mid $O_2$ | 28.9 | 13.6 | 57.2 | 0.1 | 0.2 | 2.13 | 0.74 | Cubic B1 |
| $Al_{70}Cr_{25}$ 5% Si mid $O_2$ | 31.2 | 11.5 | 56.9 | 0.2 | 0.2 | 2.71 | 0.75 | Cubic B1 |
| $Al_{70}Cr_{25}$ 5% Si high $O_2$ | 30.4 | 11.7 | 57.6 | 0.2 | 0.1 | 2.60 | 0.73 | Cubic B1 |

**Table S2.** The chemical compositions of the coatings measured with XPS (on polished Si wafers in order to have as flat surface as possible)

| Coating variants | Elemental composition [at %] | | | | | $\frac{Al}{Cr}$ | $\frac{Si*100}{Cr+Al+Si}$ | $\frac{(Al+Cr+Si)}{O}$ | Structure |
|---|---|---|---|---|---|---|---|---|---|
| | Al | Cr | Si | O | C | | | | |
| AlCr mid $O_2$ | 16.7 | 21.5 | 0 | 60.7 | 1 | 0.78 | 0 | 0.63 | Cubic B1→ Corundum |
| AlCr 5% Si low $O_2$ | 16.4 | 21.0 | 1.7 | 60.1 | 0.8 | 0.78 | 4.4 | 0.65 | Cubic B1 |
| AlCr 5% Si mid $O_2$ | 16.7 | 20.9 | 1.5 | 59.9 | 1.1 | 0.80 | 3.9 | 0.65 | Cubic B1 |
| AlCr 5% Si high $O_2$ | 16.7 | 22.0 | 1.5 | 60.1 | 0.8 | 0.76 | 3.8 | 0.65 | Cubic B1 |
| AlCr 10% Si mid $O_2$ | 16.4 | 20.0 | 3 | 59.9 | 0.8 | 0.82 | 7.6 | 0.66 | Cubic B1 |
| $Al_{70}Cr_{30}$ mid $O_2$ | 24.4 | 12.2 | 0 | 63.4 | 0 | 2.00 | 0 | 0.58 | Cubic B1 |
| $Al_{70}Cr_{25}$ 5% Si mid $O_2$ | 23.3 | 10.8 | 1.7 | 64.2 | 0 | 2.16 | 4.8 | 0.56 | Cubic B1 |
| $Al_{70}Cr_{25}$ 5% Si high $O_2$ | 23.1 | 10.8 | 1.7 | 64.4 | 0 | 2.14 | 4.7 | 0.55 | Cubic B1 |

[*] Corresponding authors: email address: ludvig.landalv@sandvik.com and Per.eklund@liu.se

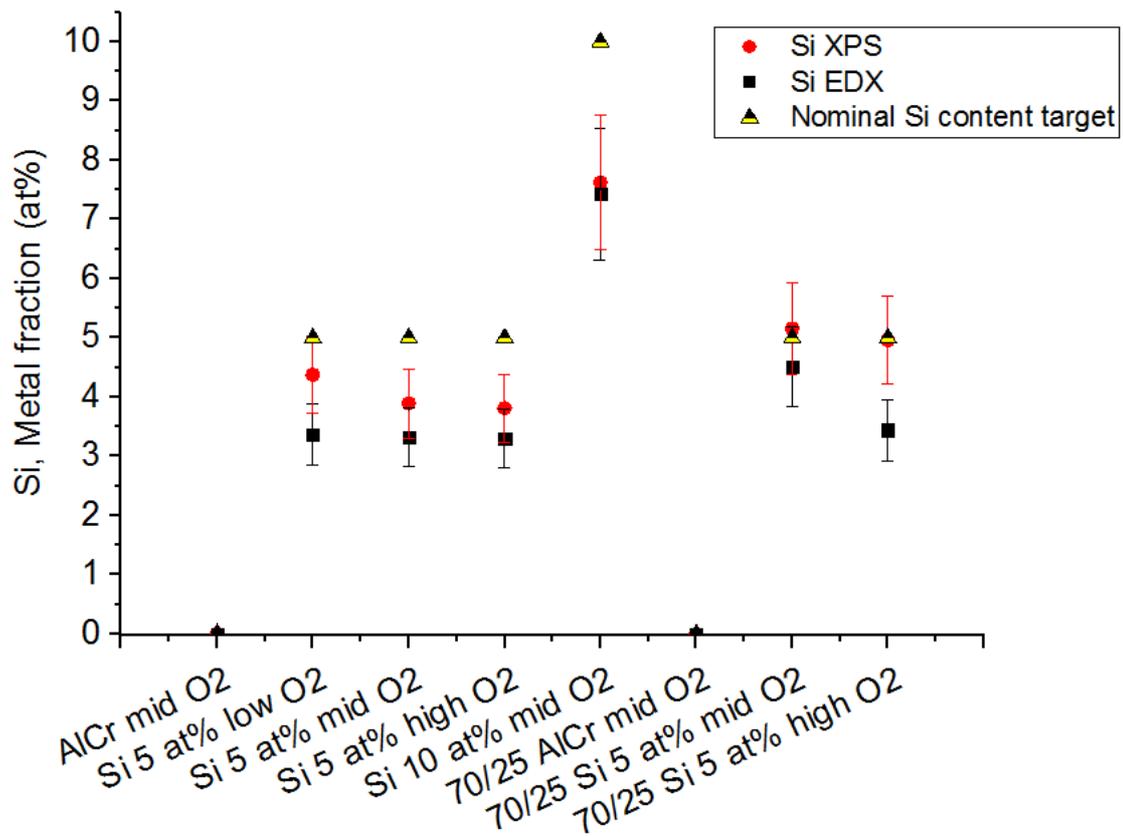

**Figure S1.** Metal fraction measured Si-content in the coating with EDX and XPS.

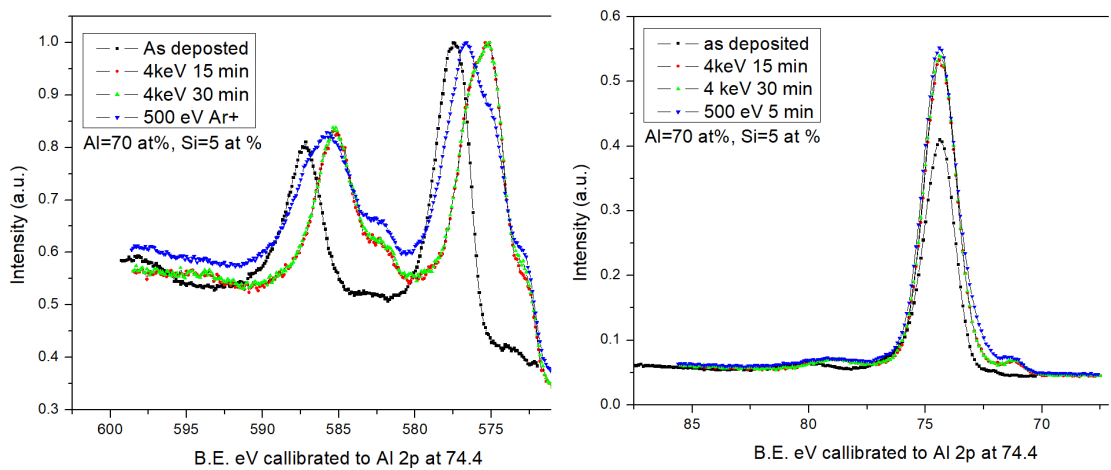

**Figure S2.** XPS spectra on as deposited and after different amount time Ar⁺ sputtering. Cr2p (left) and Al 2p (right) for the coating deposited from 70 at. % Al and 5 at. % Si cathodes. The intensity is normalized to the maximum value of the respective Cr 2p spectra and shifted to the Al 2p value of 74.4 eV.